\documentclass[aps,preprint,epsfig,rotate]{revtex4}
\begin{document}

\title{Internal structure and positron annihilation in the four-body MuPs system}
 \author{Alexei M. Frolov}
 \email[E--mail address: ]{afrolov@uwo.ca}

\affiliation{Department of Applied Mathematics\\
 University of Western Ontario, London, Ontario N6H 5B7, Canada}

\date{\today}

\begin{abstract}

A large number of bound state properties of the four-body muonium-positronium system MuPs (or $\mu^{+} e^{-}_2 e^{+}$) are 
determined to high accuracy. Based on these expectation values we predict that the weakly-bound four-body MuPs system has 
the `two-body' cluster structure Mu + Ps. The two neutral clusters Mu ($\mu^{+} e^{-}$) and Ps ($e^{+} e^{-}$) interact with
each other by the attractive van der Waals forces. By using our expectation values of the electron-positron delta-functions 
we evaluated the half-life $\tau_a$ of the MuPs system against annihilation of the electron-positron pair: $\tau_a = 
\frac{1}{\Gamma} \approx 4.071509 \cdot 10^{-10}$ $sec$. The hyperfine structure splitting of the ground state in the MuPs 
system evaluated with our expectation values is $\Delta \approx$ 23.064(5) $MHz$.

\end{abstract}
\maketitle
\newpage

In this study we consider bound state properties of the four-body muonium-positronium system MuPs, or $\mu^{+} e^{-}_2 e^{+}$. The fact 
that this system is bound is known since the middle of 1980s when it was shown that the negatively charged Mu$^{-}$ (or $\mu^{+} 
e^{-}_2$) is bound \cite{Fro1986} (calculations) and \cite{Mu-exp} (experiment). The goal of this study is to perform highly accurate 
computations of the ground bound ${}^{1}S(L = 0)-$state in the MuPs system, which is, in fact, the only  bound state in this four-body 
system. The non-relativistic Hamiltonian of the four-body $\mu^{+} e^{-}_2 e^{+}$ system is written in the form (in atomic units $\hbar 
= 1, m_e = 1, e = 1$):
\begin{eqnarray}
 H = -\frac{1}{2 m_{\mu}} \Delta_{1} -\frac{1}{2} \Delta_{2} -\frac{1}{2} \Delta_{3} -\frac{1}{2} \Delta_{4} + 
       \frac{1}{r_{12}} - \frac{1}{r_{13}} - \frac{1}{r_{14}} - \frac{1}{r_{23}} - \frac{1}{r_{24}} + 
       \frac{1}{r_{34}} \label{eq1}
\end{eqnarray}
where the notation 1 designates the positively charged muon $\mu^{+}$, the notation 2 (or +) means the positron, while 3 and 4 (or -) stand 
for electrons. The same system of notations is used everywhere below in this study. 

By solving the corresponding Schr\"{o}dinger equation $H \Psi = E \Psi$ for bound states ($E < 0$) one can determine the total energy and wave 
function of the bound (ground) $S(L = 0)-$state in the MuPs system. In general, to determine the bound state spectra in four-body MuPs system 
in this study we apply the variational expansion written in the basis of four-dimensional gaussoids, where each basis function depends upon the 
relative (or interparticle) coordinates $r_{ij} = \mid {\bf r}_i - {\bf r}_j \mid = r_{ji}$ \cite{KT} and ${\bf r}_i$ ($i$ = 1, 2, 3, 4) are 
the Cartesian coordinates of the particle $i$. Note that each of the six relative coordinates $r_{ij}$ is translationally and rotationally 
invariant. 

The expansion, Eq.(\ref{eq1}), was proposed more than 35 years ago (see, e.g., \cite{KT} and earlier references therein) to solve some nuclear and 
hypernuclear few-body problems. Since the midlle of 1980s the same expansion was also used to determine some bound states in different Coulomb 
few-body systems. In the case of four-body Coulomb systems this variational expansion takes the form
\begin{eqnarray}
 \Psi = \sum^{N}_{k=1} C_k \exp( -\alpha^{(k)}_{12} r^{2}_{12} -\alpha^{(k)}_{13} r^{2}_{13} -\alpha^{(k)}_{14} 
 r^{2}_{14} -\alpha^{(k)}_{23} r^{2}_{23} -\alpha^{(k)}_{24} r^{2}_{24} -\alpha^{(k)}_{34} r^{2}_{34}) \; \; \; , 
 \label{exp}
\end{eqnarray}
where $C_k$ are the linear variational parameters of this expansion $(k = 1, \ldots, N$), while $\alpha^{k}_{ij}$ are the non-linear parameters of 
the variational expansion, Eq.(\ref{exp}). In applications to actual four-body systems the trial wave function, Eq.(\ref{exp}), must be symmetrized
(or antisymmetrized) with respect to the possible presence of identical particles. In particular, in the MuPs system we have two identical electrons 
(particles 3 and 4). In general, the overall efficiency of the variational expansion, Eq.(\ref{exp}), depends upon algorithms which are used to optimize 
the non-linear parameters in Eq.(\ref{exp}). Recently, for four-body systems we have developed a number of algorithms which are very effective and 
produce fast optimization. The accuracy of the constructed wave functions is usually high and very high for the total energies. The expectation values of 
various geometrical and dynamical properties are also determined to relatively high numerical accuracy. However, some troubles can be found in computations 
of the expectation values of some delta-functions, cusp values and a few other similar properties.   

Results of our calculations of the ground state in the MuPs system are shown in Table I. All calculations have been performed in atomic units 
and include the total energies and some other bound state properties of this system computed for different values of $N$ in Eq.(\ref{exp}). In 
our calculations we have used the following values of $N =$ 600, 800, 1000 and 1500. The mass of the positive charged muon $\mu^{+}$ used in our 
calculations equals $m_{\mu}$ = 206.768262 $m_e$. Note that our current wave functions are significantly more accurate than analogous functions used 
in earlier studies. As follows from the results shown in Table I the internal structure of the MuPs system is represented as a two-body cluster
Mu $\longleftrightarrow$ Ps which is formed from the muonium atom Mu (or $\mu^{+} e^{-}$) and neutral positronium Ps ($e^{-} e^{+}$) weakly 
interacting with each other. In general, such an interaction of these two neutral clusters is represented by the van der Waals attracting force(s) 
which is sufficient to bind the whole four-body system together. The competing `ionic' model of MuPs fails to predict correctly many properties from 
Table I. In this ionic model the MuPs system is represented as a motion of the positron $e^{+}$ in the field of the central, heavy ion $\mu^{+} 
e^{-}_2$ which is negatively charged. In this model actual distances from the positron to the central cluster/ion are significantly larger than the 
radius of this central ion Mu$^{-}$.     

The expectation values of different operators are used to determine the properties which can later be measured in actual experiments. For instance, 
let us consider annihilation of electron-positron pair in the MuPs system. This process can be observed experimentally. It is clear that the 
largest annihilation rate corresponds to the two-photon annihilation. The formula for the two-photon annihilation width (or rate) $\Gamma_{2 
\gamma}({\rm MuPs})$ is
\begin{eqnarray}
 \Gamma_{2 \gamma}({\rm MuPs}) = 2 \pi \alpha^4 c a^{-1}_0 \Bigl[ 1 - \frac{\alpha}{\pi} \Bigl( 5 - \frac{\pi^2}{4} \Bigr)\Bigr] 
 \langle \delta({\bf r}_{+-}) \rangle = 100.34560545419 \cdot 10^{9} \langle \delta({\bf r}_{+-}) \rangle \; sec^{-1} \; \; \; . \label{An2g}
\end{eqnarray}
where $\langle \delta_{+-} \rangle$ is the expectation value of the electron-positron delta-function determined for the ground bound state in the MuPs 
system. Here and below the indexes `+' and `-' designate the positron and electron, respectively. In Eq.(\ref{An2g}) the notation $\alpha = \frac{e^2}{\hbar c} 
= 7.2973525698 \cdot 10^{-3} \Bigl(\approx \frac{1}{137}\Bigr)$ is the dimensionless fine structure constant, $c = 2.99792458 \cdot 10^{8}$ $m \cdot sec^{-1}$ 
is the speed of light in vacuum, and the Bohr radius $a_0$ equals $0.52917721092 \cdot 10^{-10}$ $m$ \cite{CRC}. Analogous formula for the three-photon 
annihilation rate $\Gamma_{3 \gamma}({\rm MuPs})$ takes the form
\begin{eqnarray}
 \Gamma_{3 \gamma}({\rm MuPs}) = 2 \frac{4 (\pi^2 - 9)}{3} \alpha^5 c a^{-1}_0 \langle \delta({\bf r}_{+-}) \rangle = 2.7185459576 \cdot
 10^8 \langle \delta({\bf r}_{+-}) \rangle \; sec^{-1}
\end{eqnarray}
In these formulas and everywhere below $\alpha = 7.2973525698 \cdot 10^{-3} \Bigl( \approx \frac{1}{137} \Bigr)$ is the dimensionless fine structure constant, 
$c = 2.99792458 \cdot 10^{8}$ $m sec^{-1}$ is the speed of light in vacuum and $a_0 = 0.52917721092 \cdot 10^{-10}$ $m$ is the Bohr radius. The numerical 
values of these constants have been taken from \cite{CRC}. 

The rates of the four- and five-photon annihilations of the electron-positron pairs in the MuPs system are uniformly related with the $\Gamma_{2 
\gamma}({\rm MuPs})$ and $\Gamma_{3 \gamma}({\rm MuPs})$ rates, respectively. The approximate relations are written in the two following forms \cite{PRA83}
\begin{equation}
 \Gamma_{4 \gamma}({\rm MuPs}) \approx 0.274 \Bigl(\frac{\alpha}{\pi}\Bigr)^2 \Gamma_{2 \gamma}({\rm MuPs})
 \approx 1.4783643 \cdot 10^{-6} \cdot \Gamma_{2 \gamma}({\rm MuPs}) \label{e4a}
\end{equation}
and
\begin{equation}
 \Gamma_{5 \gamma}({\rm MuPs}) \approx 0.177 \Bigl(\frac{\alpha}{\pi}\Bigr)^2 \Gamma_{3 \gamma}({\rm MuPs})
 \approx 9.55001778 \cdot 10^{-7} \cdot \Gamma_{3 \gamma}({\rm MuPs}) \label{e5a}
\end{equation}
By using the expectation value of the $\delta({\bf r}_{+-})-$function from Table I (2.4410195$\cdot {-2}$) we can evaluate these annihilation rates: $\Gamma_{2 \gamma}$ 
= 2.4494558$\cdot 10^{9}$ $sec^{-1}$, $\Gamma_{3 \gamma}$ = 6.6360237$\cdot 10^{6}$ $sec^{-1}$, $\Gamma_{4 \gamma}$ = 3.621188$\cdot 10^{3}$ $sec^{-1}$ and $\Gamma_{5 
\gamma}$ = 6.337414 $sec^{-1}$. Now, one can evaluate the total annihilation rate of the MuPs system by the following sum $\Gamma \approx \Gamma_{2 \gamma} + \Gamma_{3 
\gamma} + \Gamma_{4 \gamma} + \Gamma_{5 \gamma} \approx \Gamma_{2 \gamma} + \Gamma_{3 \gamma} \approx 2.4560918 \cdot 10^{9}$ $sec^{-1}$. In other words, the knowledge 
of accurate values of the $\Gamma_{2 \gamma}$ and $\Gamma_{3 \gamma}$ annihilation rates is sufficient to predict the half-life of the MuPs system against positron 
annihilation $\tau = \frac{1}{\Gamma} \approx 4.071509 \cdot 10^{-10}$ $sec$.

In addition to the few-photon annihilation discussed above in the four-body MuPs system the electron-positron pair can annihilate with the emission of one and zero 
photons. The corresponding annihilation rates are very small, but in some theoretical considerations one- and zero-photon annihilations play a noticeable role. An 
approximate formula for zero-photon annihilation rate $\Gamma_{0 \gamma}$ takes the form (see, e.g., \cite{FrWa2010}):
\begin{eqnarray}
 \Gamma_{0 \gamma} = \xi \frac{147 \sqrt{3} \pi^3}{2} \cdot \alpha^{12} (c a_0^{-1}) \cdot \langle \delta_{\mu^{+}+--} \rangle =  5.0991890 \cdot
 10^{-4} \cdot \xi \cdot \langle \delta_{\mu^{+}+--} \rangle \; \; \; sec^{-1} \label{0phot}
\end{eqnarray}
where $\langle \delta_{\mu^{+}+--} \rangle$ is the expectation value of the four-particle delta-function in the ground state of muonium-positronium (MuPs). The numerical 
value of $\langle \delta_{\mu^{+}+--} \rangle$ is the probability to find all four particles in one small volume with the spatial radius $R \approx \alpha a_0 = \Lambda_e 
(= \frac{\hbar}{m_e c})$ is the Compton wavelength. The unknown (dimensionless) factor $\xi$ has the numerical value close to unity. The expectation value of the four-particle 
delta-function determined in our calculations is $\approx 1.75854 \cdot 10^{-4}$ (in $a.u.$). From this one finds that $\Gamma_{0 \gamma}$(MuPs) $\approx 8.9671 \cdot 10^{-8} 
\xi$ $sec^{-1}$. 

One-photon annihilation rate can be evaluated by using the fact that in the lowest-order approximation the one-photon annihilation of the electron-positron pair in MuPs 
can be considered as a regular two-photon annihilation, but one of the two emitted photons is absorbed either by the remaining electron $e^{-}$, or by the muon
$\mu^{+}$. This leads to the two different one-photon annihilation rates which are designated below as $\Gamma^{(1)}_{1 \gamma}$ and $\Gamma^{(2)}_{1 \gamma}$, respectively.
In the case of absorbtion by an electron the probability of this process is given by the formula
\begin{eqnarray}
 \Gamma^{(1)}_{1 \gamma} = \frac{64 \pi^2}{27} \cdot \alpha^{8} (c a_0^{-1}) \cdot \langle \delta_{+--} \rangle = 1.065756921658 \cdot 10^3
 \cdot \langle \delta_{+--} \rangle \; \; \; sec^{-1} \; \; \; , \; \; \; \label{1gamma}
\end{eqnarray}
where $\langle \delta_{+--} \rangle$ is the expectation value of the triple electron-positron-electron delta-function determined for the ground state of the MuPs system. 
Its numerical value is the probability to find all three particles inside of a sphere  which has spatial radius $R \approx \alpha a_0 \approx \frac{a_0}{137} = \Lambda_e$. Our 
best numerical treatment to-date for the $\langle \delta_{+--} \rangle$ value gives $\approx 3.65954 \cdot 10^{-4}$, and therefore, $\Gamma^{(1)}_{1 \gamma} \approx$ 
3.90018$\cdot 10^{-1}$ $sec^{-1}$ for the bound (ground) state in the MuPs system. 

Analysis of the second one-photon annihilation of the $(e^{+},e^{-})-$pair in the MuPs system is more complicated (see discussion in \cite{FrWa2010}). An approximate 
expression for the $\Gamma^{(2)}_{1 \gamma}$ rate is written in the form which is similar to Eq.(\ref{1gamma})
\begin{eqnarray}
 \Gamma^{(2)}_{1 \gamma} = \chi \frac{64 \pi^2}{27} \cdot \alpha^{8} (c a_0^{-1}) \cdot \langle \delta_{\mu^{+}+-} \rangle = 1.065756921658 \cdot 10^3 \cdot \langle 
 \delta_{\mu^{+}+-} \rangle \; \; \; sec^{-1} \; \; \; , \; \; \; \label{1gamma2}
\end{eqnarray}
where $\langle \delta_{\mu^{+}+-} \rangle$ is the expectation value of the triple muon-electron-positron delta-function determined for the ground state of the MuPs system
and factor $\chi$ is a numerical factor which is approximately equal to the factor $\xi$ in Eq.(\ref{0phot}). To produce more accurate formulas for $\Gamma^{(2)}_{1 \gamma}$ 
and exact expressions for the factors $\xi$ in Eqs.(\ref{0phot}) and $\chi$ in Eq.(\ref{1gamma2}) one needs to perform an adittional analysis. 

For the muonium-positronium system MuPs there is a possibility to observe an interesting process which is called the muon-positron conversion. In general, the muon 
decay is written in the form $\mu^{+} = e^{+} + \nu_e + \overline{\nu}_{\mu}$, where the notation $\nu_e$ stands for the electron neutrino, while the notation 
$\overline{\nu}_{\mu}$ designates the muonic anti-neutrino. The decay equation for the positively charged muon re-written from the left-to-right represents a creation (or 
synthesis) of the $\mu^{+}$ muon, i.e. $e^{+} + \nu_e + \overline{\nu}_{\mu} = \mu^{+}$. Since the MuPs system already contains a positron $e^{+}$, then these two processes 
(muonic decay and muon synthesis) can proceed instantly and we can observe a `self-transition' of MuPs into MuPs. In our earlier paper \cite{Fro2012} we have evaluated the 
probability to observe the muon-positron conversion as one event for $\approx 1 \cdot 10^{8}$ of MuPs systems. This means that currently we cannot observe the muon-positron 
conversion in MuPs, since the probability of conversion is extremely small and it is still very difficult to create even one MuPs system. However, in the future this 
situation can be changed and one can study the muon-positron conversion in the MuPs system experimentally.

In conclusion, let us determine the hyperfine structure splitting in the MuPs system. Such a structure arises from interaction between the spin-vectors of the positron and 
muon. The hyperfine structure splitting in the MuPs system is written in the form 
\begin{equation}
 a = \frac{8 \pi \alpha^2}{3} \mu^2_B \frac{g_{\mu}}{m_{\mu}} \frac{g_{e}}{m_{e}} \cdot \langle \delta_{\mu^{+} e^{+}} \rangle = 14229.1255 \cdot 
 \langle \delta_{\mu^{+} e^{+}} \rangle \label{spl31}
\end{equation}
where $\alpha$ is the fine structure constant, $\mu_B$ is the Bohr magneton which is exactly equal 0.5 in atomic units, while $\langle \delta_{\mu^{+} e^{+}} \rangle$ is the 
expectation value of the muon-positron delta-function. Also in Eq.(\ref{spl31}) the notations $m_{\mu}$ and $m_{e}$ stand for the mass-at-rest for the muon and positron,
respectively, while the factors $g_{+}$ = -2.0023193043718 and $g_{\mu}$ = -2.0023318396 are the gyromagnetic ratios. By using the expectation value of the muon-positron 
delta-function $\langle \delta_{\mu^{+} e^{+}} \rangle \approx 1.620893 \cdot 10^{-3}$ $a.u.$ from Table I, one finds that the value $a$ in Eq.(\ref{spl31}) equals $a 
\approx 23.064$ $MHz$. This coincides with the energy difference between the hyperfine structure states with $J = 0$ and $J = 1$, where the notation $J$ stands for the total 
spin of the muon-positron pair in the MuPs system.

We have considered the bound state properties of the MuPs system ($\mu^{+} e^{-}_2 e^{+}$, or muonium-positronium). As follows from our computational results of bound state 
properties the internal structure of the MuPs system is represented to very good accuracy as a two-body cluster Mu + Ps. The two neutral systems Mu ($\mu^{+} e^{-}$) and Ps 
($e^{+} e^{-}$) interact with each other by the attractive van der Waals forces. By using our results from accurate computations we determine a few annihlation rates of the 
electron-positron pair in the MuPs system. Numerical values of the two-, three-, four- and five-photon annihilation rates of the MuPs system are determined to high numerical 
accuracy. The rate of zero-photon annihilation $\Gamma_{0 \gamma}$(MuPs) and first one-photon annihilation rate $\Gamma^{(1)}_{1 \gamma}$(MuPs) have been evaluated 
approximately. Another interesting property which we also determine in this study is the hyperfine structure splitting between singlet $J = 0$ and triplet $J = 1$ spin states 
of the muon-positron pair in MuPs. By using our expectation value of the $\mu^{+} - e^{+}$ delta-function we have found that the hyperfine splitting $\Delta$ in the ground 
state of the MuPs system is $\approx$ 23.064(5) $MHz$. The values ($\Gamma_{2 \gamma}, \Gamma_{3 \gamma}, \Gamma$ and $\Delta$) determined in this study can directly be measured 
in future experiments. 

Results of our study indicate clearly that many bound state properties, including properties which can be measured in modern experiments, have now been determined to good 
numerical accuracy. The next step is to perform an experiment to create the actual MuPs system, observe its decay and measure some of the properties. At this moment we can predict 
that further changes in theoretically determined values will be small and even negligible. On the other hand, it is clear that without actual experiments we cannot make any visible 
progress in this area of research. 

\begin{center}
       {\bf Acknowledgments}
\end{center}

I am grateful to David M. Wardlaw from the Memorial University of Newfoundland (St.John's, Newfoundland, CANADA) for helpful discussions and inspiration. 

%

 \begin{table}[tbp]
   \caption{The expectation values of a number of bound state properties in atomic units of the MuPs system ($\mu^{+} e^{-}_2 e^{+}$). The notation $\mu$
            designates the positively charged muon, while the notations $e^{-}$ and $e^{+}$ denote the electron and positron, respectively.} 
     \begin{center}
     \begin{tabular}{| c | c | c | c | c | c | c | c |}
       \hline\hline          
 $N$ & $E$ & $\langle r^{-2}_{\mu - e^{+}} \rangle$ &  $\langle r^{-1}_{\mu - e^{+}} \rangle$ & $\langle r_{\mu - e^{+}} \rangle$ & $\langle r^{2}_{\mu - e^{+}} \rangle$ & $\langle r^{3}_{\mu - e^{+}} \rangle$ & $\langle r^{4}_{\mu - e^{+}} \rangle$ \\
      \hline
 600 & -0.78631701225 & 0.1708476 & 0.3460479 & 3.678243 & 16.41074 & 86.4005 & 527.295 \\

 800 & -0.78631707338 & 0.1708477 & 0.3460481 & 3.678240 & 16.41068 & 86.3994 & 527.272 \\

1000 & -0.78631708004 & 0.1708476 & 0.3460482 & 3.678237 & 16.41065 & 86.3986 & 527.257 \\

1500 & -0.78631708035 & 0.1708476 & 0.3460483 & 3.678236 & 16.41063 & 86.3984 & 527.251 \\
     \hline\hline
 $N$ & $\langle \frac12 p^2_{e^{-}} \rangle$ & $\langle r^{-2}_{e^{-} - e^{+}} \rangle$ &  $\langle r^{-1}_{e^{-} - e^{+}} \rangle$ & $\langle r_{e^{-} - e^{+}} \rangle$ & $\langle r^{2}_{e^{-} - e^{+}} \rangle$ & $\langle r^{3}_{e^{-} - e^{+}} \rangle$ & $\langle r^{4}_{e^{-} - e^{+}} \rangle$ \\
       \hline
 600 & 0.32338448362  & 0.3485124 & 0.4178999 & 3.4882569 & 15.666223 & 85.1136 & 539.805 \\ 

 800 & 0.32338468807  & 0.3485146 & 0.4179002 & 3.4882549 & 15.666187 & 85.1129 & 539.790 \\

1000 & 0.32338474500  & 0.3485150 & 0.4179003 & 3.4882533 & 15.666160 & 85.1124 & 539.781 \\

1500 & 0.32338474874  & 0.3485151 & 0.4179005 & 3.4882530 & 15.666155 & 85.1121 & 539.778 \\
     \hline\hline
 $N$ & $\langle \frac12 p^2_{e^{+}} \rangle$ & $\langle r^{-2}_{e^{-} - \mu} \rangle$ &  $\langle r^{-1}_{e^{-} - \mu} \rangle$ & $\langle r_{e^{-} - \mu} \rangle$ & $\langle r^{2}_{e^{-} - \mu} \rangle$ & $\langle r^{3}_{e^{-} - \mu} \rangle$ & $\langle r^{4}_{e^{-} - \mu} \rangle$ \\
       \hline
 600 & 0.13668660960  & 1.1945831 & 0.7257294 & 2.3260105 & 7.9172577 & 35.9559 & 204.728 \\ 

 800 & 0.13668676776  & 1.1945864 & 0.7257297 & 2.3260092 & 7.9172364 & 35.9555 & 204.719 \\

1000 & 0.13668681862  & 1.1945869 & 0.7257298 & 2.3260078 & 7.9172134 & 35.9551 & 204.712 \\

1500 & 0.13668681473  & 1.1945871 & 0.7257299 & 2.3260075 & 7.9172127 & 35.9550 & 204.708 \\
     \hline\hline
 $N$ & $\langle \frac12 p^2_{\mu} \rangle$ & $\langle r^{-2}_{e^{-} - e^{-}} \rangle$ &  $\langle r^{-1}_{e^{-} - e^{-}} \rangle$ & $\langle r_{e^{-} - e^{-}} \rangle$ & $\langle r^{2}_{e^{-} - e^{-}} \rangle$ & $\langle r^{3}_{e^{-} - e^{-}} \rangle$ & $\langle r^{4}_{e^{-} - e^{-}} \rangle$ \\
       \hline
 600 & 0.59169507317  & 0.2115961 & 0.3685766 & 3.5945579 & 16.056830 & 86.0545 & 540.977 \\

 800 & 0.59169544573  & 0.2115953 & 0.3685768 & 3.5945554 & 16.056785 & 86.0536 & 540.958 \\  

1000 & 0.59169564501  & 0.2115954 & 0.3685769 & 3.5945526 & 16.056739 & 86.0529 & 540.944 \\ 

1500 & 0.59169557386  & 0.2115955 & 0.3685770 & 3.5945520 & 16.056731 & 86.0523 & 540.940 \\
     \hline\hline

 $N$ & $\langle \delta({\bf r}_{e^{-} - e^{+}}) \rangle$ & $\langle \delta({\bf r}_{\mu - e^{+}}) \rangle$ & $\langle \delta({\bf r}_{\mu - e^{-}}) \rangle$ & $\langle \delta({\bf r}_{\mu e^{-} e^{+}}) \rangle$ & 
       $\langle \delta({\bf r}_{e^{-} e^{-} e^{+}}) \rangle$ & $\langle \delta({\bf r}_{\mu e^{-} e^{-}}) \rangle$ & $\langle \delta({\bf r}_{\mu e^{-} e^{-} e^{+}}) \rangle$ \\
       \hline
 600 &  0.024401081 & 0.001624112 & 0.17419597 & 8.59175$\cdot 10^{-4}$ & 3.67156$\cdot 10^{-4}$ & 7.1599$\cdot 10^{-3}$ & 1.75638$\cdot 10^{-4}$ \\

 800 &  0.024410140 & 0.001623304 & 0.17426468 & 8.58962$\cdot 10^{-4}$ & 3.66187$\cdot 10^{-4}$ & 7.2092$\cdot 10^{-3}$ & 1.77569$\cdot 10^{-4}$ \\

1000 &  0.024410167 & 0.001620911 & 0.17426917 & 8.58744$\cdot 10^{-4}$ & 3.66211$\cdot 10^{-4}$ & 7.1895$\cdot 10^{-3}$ & 1.76209$\cdot 10^{-4}$ \\

1500 &  0.024410195 & 0.001620893 & 0.17426911 & 8.56456$\cdot 10^{-4}$ & 3.65954$\cdot 10^{-4}$ & 7.1890$\cdot 10^{-3}$ & 1.75854$\cdot 10^{-4}$ \\
     \hline\hline
  \end{tabular}
  \end{center}
   \end{table}
\end{document}